\input harvmac

\overfullrule=0pt
\parindent=0pt


\def\NSNS{{$NS\otimes NS$}}
\def\RR{{$R\otimes R$}}

\def\calZ{{\cal Z}}
\def\calP{{\cal P}}
\def\calG{{\cal G}}


\def\xxx#1 {{hep-th/#1}}
\def\lr { \lref}
\def\npb#1(#2)#3 { Nucl. Phys. {\bf B#1} (#2) #3}
\def\rep#1(#2)#3 { Phys. Rept.{\bf #1} (#2) #3}
\def\plb#1(#2)#3{Phys. Lett. {\bf #1B} (#2) #3}
\def\prl#1(#2)#3{Phys. Rev. Lett.{\bf #1} (#2) #3}
\def\physrev#1(#2)#3{Phys. Rev. {\bf D#1} (#2) #3}
\def\ap#1(#2)#3{Ann. Phys. {\bf #1} (#2) #3}
\def\rmp#1(#2)#3{Rev. Mod. Phys. {\bf #1} (#2) #3}
\def\cmp#1(#2)#3{Comm. Math. Phys. {\bf #1} (#2) #3}
\def\mpl#1(#2)#3{Mod. Phys. Lett. {\bf #1} (#2) #3}
\def\ijmp#1(#2)#3{Int. J. Mod. Phys. {\bf A#1} (#2) #3}

\def\tr{{\rm tr}}

\def\lam16{\lambda^{16}}

\parindent 25pt
\overfullrule=0pt
\tolerance=10000



\lr\wittb{E.~Witten, {\it String theory dynamics in various dimensions},
\npb443(1995)85, \xxx9503124.}

\lr\polchinst{J. Polchinski, {\it Combinatorics of boundaries in string theory}, 
\xxx9407031, \physrev50(1994)6041.} 
\lr\greeninst{M.B.  Green, {\it A gas of D-instantons}, \xxx9504108,  \plb354(1995) 271  .}  
\lr\gv{M.B.~Green and P.~Vanhove, {\it D-Instantons, strings and M
theory},
\plb408(1997)122, \xxx9704145.}

\lr\wittentop{E.~Witten, {\it Mirror Manifolds And Topological Field Theory}, 
\xxx9112056.}
\lr\narain{ I.~Antoniadis, E.~Gava, K.S.~Narain and  
T.R.~Taylor, {\it N=2 Type 
II- Heterotic duality and Higher derivative F-terms}, \xxx9507115, 
\npb455(1995)109}
\lr\japanese{N.~Ishibashi, H.~Kawai, Y.~Kitazawa and A.~Tsuchiya, {\it A large
 N 
reduced model as superstring}, \xxx9612115, \npb498(1997)467.}
\lr\greengutb{M.B.~Green and M.~Gutperle, $\;${\it Configurations of two
D-instantons},
 \xxx9612127, $\;$ \plb398(1997)69.}
\lr\yia{P. Yi, {\it Witten index and threshold bound states of
D-branes},
\xxx9704098, \npb505(1997)307. }
\lr\sethia{S.~Sethi and M.~Stern, {\it D-brane bound states redux},
\xxx9705046.
}
\lr\greengutc{M.B.~Green and M.~Gutperle, {\it D-particle bound states and the 
D-instanton measure}, \xxx9711107,$\;$  {\bf JHEP01}(1998)005.}
\lr\greengutd{M.B.~Green and M.~Gutperle, {\it Effects of D-instantons}, 
\xxx9701093, 
\npb498(1997)195.}
\lr\greengute{M.B.~Green,  M.~Gutperle and H.~Kwon, {\it Sixteen fermion
and related
terms in M theory on $T^2$}, \xxx9710151.}
\lr\mooreetal{G.~Moore, N.~Nekrasov and  S.~Shatahvilli, {\it D-particle bound 
states and generalized instantons}, \xxx9803265.}
\lr\nicolai{W.~Krauth, H.~Nicoai and M.~Staudacher, {\it Monte Carlo approach 
to M-theory}, \xxx9803117.}
\lr\berkov{N. Berkovits, {\it Construction of $R^4$ terms in N=2 D=8
superspace}, \xxx9709116.}
\lr\thooft{G. 't Hooft, {A property of electric and magnetic flux in nonabelian 
gauge theories},   Nucl. Phys. {\bf B153}  (1979) 141.}
\lr\aspinwall{P.~Aspinwall, {\it Some Relationships between Dualities in
 String 
Theory}, \xxx9508154 ,   Nucl.Phys.Proc.Suppl. {\bf 46} (1996) 30.}
\lr\schwarz{J.H.~Schwarz, {\it An $SL(2,Z)$ Multiplet of Type 
IIB Superstrings}, 
\xxx9508143}
\lr\antoniadis{I.~Antoniadis, B.~Pioline and T.R.~Taylor, {\it Calculable
$e^{-1/\lambda}$ effects}, \xxx9707222, \npb512(1998)61.}
\lr\piolinea{B.~Pioline, {\it A note on non-perturbative $R^4$ couplings}, \xxx9804023}
\lr\stromingera{A.~Strominger, {\it Massless Black Holes and Conifolds in 
String 
Theory}, \xxx9504090, \npb451(1995)109.}
\lr\becker{K.~Becker, M.~Becker and  A.~Strominger, {\it Fivebranes, Membranes 
and Non-Perturbative String Theory}, \xxx9507158, \npb456(195)130.}
\lr\greene{B.R.~Greene, D.R.~Morrison and C.~Vafa, {\it A geometric
 realization 
of confinement}, \xxx9608039, \npb481(1996)513.}
\lr\ooguri{H.~Ooguri and C.~Vafa, {\it Summing up D-Instantons},  
\xxx9608079,$\;$\prl77(1996)3296.}
\lr\oogurioz{ H.~Ooguri, Y.~Oz and Z.~Yin, {\it D-Branes on Calabi-Yau Spaces 
and Their Mirrors}, \xxx9606112, \npb477(1996)407. }
\lr\guta{M.~Gutperle, {\it Aspects of D-Instantons}, \xxx9712156.}
\lr\dualhet{D.~L\"ust, {\it String Vacua with N=2 Supersymmetry in Four 
Dimensions}, \xxx9803072.}
\lr\porrati{M.~Porrati and A.~Rozenberg, {\it Bound states at
threshold in
supersymmetric quantum  mechanics}, \xxx9708119.}
\lr\beckerc{K.~Becker, M.~Becker, D.R.~Morrison, H.~Ooguri, Y.~Oz and  Z.~Yin,
 {\it Supersymmetric Cycles in Exceptional Holonomy Manifolds and Calabi-Yau
     4-Folds}, \xxx9608116, \npb480(1996)225.}
\lr\shenkerseiberg{S.~Shenker and N.~Seiberg, {\it Hypermultiplet Moduli Space 
and String Compactification to Three Dimensions}, \xxx9608086, 
\plb388(1996)521.}


\lr\bacha{C.~Bachas, C.~Fabre, E.~Kiritsis, N.A.~Obers and  P.~Vanhove,
 {\it Heterotic / type I duality and D-brane instantons}, \xxx9707126,
 \npb509(1998)33.}

\lr\bachb{C.~Bachas, {\it  Heterotic versus Type I}, \xxx9710102,$\;$ Talk at 
STRINGS'97.}

\lr\kiritsisa{E.~Kiritsis and  N.A.~Obers, {\it  Heterotic/Type-I Duality in
 D$<$10 Dimensions, Threshold Corrections and
     D-Instantons}, \xxx9709058.}

\lr\kiritsisb{E.~Kiritsis and  B.~Pioline, {\it On $R^4$ threshold
 corrections in IIB string theory and (p,q) string instantons}, 
\xxx9707018, \npb508(1997)509.}
        
\lr\kiritsisc{E.~Kiritsis and B.~Pioline, {\it U-duality and
 D-brane Combinatorics},$\;$\xxx9710078, \plb418(1998)61.}

\lr\piolinea{B.~Pioline, {\it A note on non-perturbative $R^4$ couplings},
\xxx9804023.}
\lr\russoa{J. Russo and  A.A. Tseytlin, {\it One-loop four-graviton
 amplitude in eleven-dimensional supergravity}, \xxx970713, \npb508(1997)245.}

\lr\russob{J.G. Russo, {\it An ansatz for a non-perturbative 
four-graviton amplitude in type IIB superstring
     theory}, \xxx9707241.} 

\lr\russoc{J.G. Russo, {\it Construction of SL(2,Z) invariant
 amplitudes in type IIB superstring theory}, \xxx9802090.}

\lr\kehaa{A. Kehagias and  H. Partouche, {\it D-Instanton Corrections as 
(p,q)-String Effects and Non-Renormalization Theorems}, \xxx9712164.}

\lr\kehab{A. Kehagias and  H. Partouche, {\it On the exact quartic 
effective action for the type IIB superstring}, \xxx9710023.}

\lr\berkova{N. Berkovits and C. Vafa, {\it Type IIB $R^4 H^{4g-4}$
 Conjectures}, \xxx9803145.}

\lr\lustb{K. Behrndt, I. Gaida, D. Lust, S. Mahapatra and  T. Mohaupt,
 {\it From Type IIA Black Holes to T-dual Type IIB D-Instantons in N=2, D=4
     Supergravity}, \xxx9706096, \npb508(1997)659.}
\lr\bergb{ E. Bergshoeff and  K. Behrndt, {\it D-Instantons and asymptotic
 geometries}, \xxx9803090.}

 \lr\eguchi{T.  Eguchi and H. Kawai, {\it Reduction of dynamical degrees of 
freedom in the large $N$ gauge theory},    Phys. Rev. Lett. {\bf 48} (1982) 
1063.}
 \lr\periwal{V.  Periwal, {\it Matrices on a point as the theory of 
everything}, \xxx9611103, \physrev55(1997)1711.}

\lr\wittens{E.  Witten, {\it Some comments on string dynamics},  in Strings95: 
Future Perspectives in String Theory, Los Angeles, CA, 13-18 Mar
1995,  xxx9507121.} 


\noblackbox
\baselineskip 14pt plus 2pt minus 2pt
\Title{\vbox{\baselineskip12pt
\hbox{hep-th/9804123}
\hbox{DAMTP-98-22}
\hbox{PUPT-1789}
\hbox{NSF-ITP-98-051}
}}
{\vbox{
\centerline{D-instanton partition functions }
  }}
\centerline{ Michael B. Green,}
\medskip
\centerline{Institute for Theoretical Physics, Santa Barbara, CA  93106-4030, 
USA }
\centerline{and}
\centerline{DAMTP, Silver Street, Cambridge CB3 9EW, UK} \centerline{\it
M.B.Green@damtp.cam.ac.uk} \bigskip
\centerline{Michael Gutperle}
\medskip
\centerline{Joseph Henry Laboratories, Princeton University}
\centerline{Princeton, New Jersey 08544,  USA}
 \centerline{\it gutperle@feynman.princeton.edu}

\bigskip
 \medskip

\centerline{{\bf Abstract}}

Duality arguments are used to determine  D-instanton 
contributions to 
certain effective interaction terms of type II 
supergravity theories in various dimensions.  This leads to
exact expressions for
the partition functions of the finite $N$ D-instanton matrix model in $d=4$ 
and $6$ dimensions that generalize our previous expression for the 
case  $d=10$.  These results are consistent with the fact that the  Witten 
index of the T-dual D-particle process should only
 be non-vanishing for $d=10$.    

\Date{April 1998}
      

\noblackbox
\baselineskip 14pt plus 2pt minus 2pt

\newsec{Introduction}

When Yang--Mills theory with gauge group $G$ 
is reduced to zero space-time dimensions it is a theory of vector potentials 
$A_{\mu}$ (where $\mu = 0,1, \cdots d-1$ labels $d$ euclidean dimensions)
  which 
are constant matrices and the
  action is simply proportional to  $\tr [A_\mu,A_\nu]^2$.    This system 
is of 
relevance in 
  calculating the zero-mode contribution  to Yang--Mills theory in a box 
  \thooft. When $G$ is $SU(N)$ (so that $A_\mu$ is a hermitian traceless 
$N\times N$ matrix )
   and in the limit $N \to \infty$ 
 this zero-dimensional model was shown 
 by Eguchi and Kawai to encode the information
 in $d$-dimensional Yang--Mills theory \eguchi, at least in the quenched version 
of the model.   
   
Adding fermions in a supersymmetric manner dramatically changes the 
nature of the model.  In supersymmetric $SU(N)$ Yang-Mills theory, which
 exists 
in 
$d=3,4,6,10$, the fermions, $\psi^a$,
 are space-time 
spinors ($a=1, \cdots, 2^{d/2-1}$ for $d\ne 3$ and $a=1,2$ for $d=3$)
and are also 
$N\times N$ matrices\foot{We will define the fermions with Minkowski signature 
for space-time before passing to euclidean signature.}   In these cases the 
partition function can be 
written as
\eqn\partdef{Z_N^{(d)} = {1\over {\rm Vol} (SU(N))}\int DA\; D\psi \;
e^{-{1\over g}S_{YM}[A,\psi]} 
,}
where $S_{YM}$ is the supersymmetric Yang-Mills action in $d$ dimensions 
reduced to a point,
\eqn\actdeff{S_{YM} = {1\over 4}\tr\big( [A_\mu,  A_\nu]^2\big) 
+ {i\over 2}\tr 
\big(\bar \psi 
\Gamma_\mu [A^\mu, 
\psi]\big),}
and $g$ is the string coupling constant.  The large-$N$ 
limit of the $d=10$ model has been used to define a version
of the matrix theory \japanese\periwal\ that is a candidate model for a
 non-perturbative
description of IIB superstring theory.   An important  effect of the 
supersymmetry is
to ensure that there are exactly flat directions along which the
 eigenvalues of 
$A_\mu^I$ feel no potential (where $I$ label elements of the Cartan subalgebra 
of $SU(N)$).
  This can be
seen already in the $N=2$ (2 D-instanton) system where the behavior of the
\lq quenched' system, in which the instantons are separated by a fixed distance
$L$, is qualitatively different for the supersymmetric cases \greengutb.

The partition function $Z_N^{(d)}$ is also of interest in a separate context.
It may be identified with the bulk contribution
to the Witten index for the system of $N$ interacting D-particles in the 
T-dual IIA string  theory.  Indeed, it was in the context of the 
Witten index that the partition function was evaluated explicitly 
 for the case $N=2$ in \yia\sethia\  
where it was shown that  $Z^{(3)}_2 = 0$, $Z^{(4)}_2 =  Z^{(6)}_2=
1/4$ and $Z^{(10)}_2 = 5/4$.   
 These values are in accord with lore 
 concerning the presence of D-particle threshold
 bound states in $d=10$ Yang--Mills 
 matrix quantum mechanics and the absence of bound states in the 
lower-dimensional
 cases.  
This lore is based on a variety of duality arguments.  D-particles in $d=10$
are supposed to be  identified with Kaluza--Klein modes of eleven-dimensional 
supergravity and the presence of 
threshold bound states corresponds to
the multiply-charged
modes.  One signal for these states is that the Witten index for 
$N$ D-particles
should equal one.
The $d=6$ supersymmetric Yang--Mills matrix model can be obtained as
a limit of type IIA string theory compactified on K3.  This is a scaling limit 
in which a
two-cycle in the K3 vanishes.
  A  D2-brane wrapped around the 
cycle is massless at the degeneration point and is interpreted as a 
Yang--Mills gauge particle in the dual heterotic picture \wittens. 
 Since there is a
single Yang--Mills state it is important that multiply-wrapped D2-branes do
not give rise to new threshold bound states and the Witten index should vanish.
  A similar argument applies to 
the four-dimensional theory obtained by compactifying type IIB string
theory on a Calabi--Yau space in the limit in which a three-cycle is 
degenerating.  As shown by Strominger \stromingera\  the singularity in the 
classical 
vector moduli space is resolved by the presence of new massless states 
associated with a D3-brane  wrapped once around the cycle.  Again, multiple
wrappings must not give extra normalizable states if the mechanism for
resolving the singularity is to work. 
The $d=3$ partition function $Z^{(3)}_N$ is believed to vanish 
\sethia\yia\nicolai\  
so this case is qualitatively 
different from the higher-$d$ cases.  The relevant euclidean brane
configuration 
would be the euclidean D2-brane wrapping a 
three-cycle in a Calabi-Yau fourfold.
 However, such cycles do not preserve half the 
supersymmetries \beckerc. 

In \greengutc\
we demonstrated how the $d=10$ zero-dimensional matrix model
partition function for arbitrary $N$, $Z^{(10)}_N$,  
could be extracted from the expressions for certain protected terms
in the IIB effective action (such as the $R^4$ term).  We also showed how this 
is compatible with a non-vanishing Witten index.   
This discussion  will be reviewed and extended in section 2. 
The $d=6$ case will be considered in section 3 where the  matrix model  
of relevance close to a degeneration point 
at which a single two-cycle  in $K3$ shrinks to zero volume will be arrived at 
by considering a scaling limit of the 
threshold corrections derived in \antoniadis.
  In section 4 this will be further extended to the
$d=4$ case where the matrix
model is associated with the degeneration of three-cycles in
Calabi--Yau spaces of \stromingera.

In all these cases we are interested in extracting the leading instanton 
contributions to the various terms in the effective action.  These terms
 can be 
expressed in the string frame in the form,
\eqn\gendef{\int d^d x \sqrt g F_{\cal P}^{(d)}(\rho, \bar \rho)
 {\cal P}(\Psi)  
+ c.c.,}
where ${\cal P}(\Psi)$ is a combination of fields (such as $R^4$ and related
terms in the $d=10$ 
case) and $\rho = \chi + i e^{-\phi}$ is a complex combination of 
\RR\ and \NSNS\ scalar fields.  The particular moduli that are identified as 
$\chi$ 
and $\phi$ depend on the context.
For large 
$e^{-\phi}$ (weak coupling) 
the function $F_{\cal P}^{(d)}(\rho,\bar \rho)$ can be written as an expansion 
that 
has the generic form
\eqn\genexp{F_{\cal P}^{(d)} = {\rm pert.} + 
\sum_{N\ne 0}\calG_{N,{\cal P}}^{(d)}(\rho,\bar \rho),}
where ${\rm pert}$. denotes a finite number of perturbative contributions and  
$\calG_{N,{\cal P}}^{(d)}$ contains $N$-instanton effects.  To leading  
order in 
the  
weak coupling expansion   this has the form
\eqn\coefform{\eqalign{\calG_{N, {\cal P}}^{(d)} & \sim  c Z^{(d)}_N 
 (S_N)^{a_d +p} e^{-2\pi  
(S_N + iN \chi)}(1 + o(1/Ne^{-\phi})) \cr
& \equiv \calZ^{(d)}_N S_N^p,  \cr} }
where the constant  $a_d$ depends only on the dimension $d$ while  $p$ is the 
number of fields in the interaction $\calP$ in linearized approximation and 
$S_N$ is the $N$-instanton
action, $S_N = Ne^{-\phi}$.  The 
quantity  $Z_N^{(d)}$,  is identified with  the partition function 
defined 
earlier in terms of the 
zero-dimensional reduction of super Yang--Mills.  The  measure
 $\calZ^{(d)}_N = (S_N)^{a_d}e^{-2\pi (S_N + iN\chi)}$, defined by
 \coefform,   is independent of which 
interaction term is being considered.
The general structure of these interactions  can be obtained by considering 
perturbative string contributions  around a D-instanton background.  This 
requires a sum  over world-sheets that includes 
configurations of  disconnected disks with Dirichlet boundary conditions, as 
described in \greengutd\ for the $d=10$ case (and in \polchinst\greeninst\ for 
the 
bosonic string).  The lowest non-trivial perturbative contributions of this 
kind 
will be evaluated for the $d=4$ case in section 4.1  in order to confirm the 
general structure 
of 
these corrections, although  perturbation theory alone cannot verify the 
detailed form of $Z_N^{(d)}$.

We will discover that
\eqn\mures{Z_N^{(10)} = \sum_{m|N} {1\over m^2}, \qquad
 Z_N^{(6)} =Z_N^{(4)}= {1\over N^2} 
.}
The $d=10$ result was given in \greengutc\ and agrees with
 the expectation that 
the Witten index should equal one.  The
  $d=6$ and $d=4$ results agree with the expectation that the Witten index 
vanishes in these cases as well as with the results of \mooreetal.   
The relation of these results to the arguments in \porrati\ is obscure.

\newsec{$d=10$ and M-theory  - type IIB duality.}

An important ingredient of the web of non-perturbative dualities is the 
correspondence between M-theory on $T^2$ and type II string theory on 
$S^1$.  The original arguments for this were motivated by the form of the 
leading terms in the low energy effective actions for these theories 
\wittb\aspinwall\schwarz.   This generalizes to the richer structure of
 the next 
terms in the low energy expansion that include, for example, the $R^4$
 terms and 
other terms of the same dimension.  We will review the form of these terms in 
this section  and how they are related to the Witten index for the $d=10$ 
D-particle system.   

The connection with the Witten index was made in \greengutc\ where we 
suggested 
 how the analysis  of the  two D-particle system in \yia\ and \sethia\ should 
generalize to an 
arbitrary number of D-particles in  the
case of $d=10$.  
For any value of $d$ the index has the form 
\eqn\indform{I^{(N)} = \lim_{\beta \to \infty} 
\tr (-1)^F e^{-\beta H_N} \equiv I_{bulk}^{(N)} + \delta I^{(N)},}
where $H_N$ is the $N$ D-particle hamiltonian.
 $I^{(N)}_{bulk}$ is the bulk contribution to the index and is identical
to the zero-dimensional partition function, 
\eqn\bulkdef{I^{(N)}_{bulk} = \lim_{\beta\to 0}\tr (-1)^F e^{-\beta H_N} =  
Z_N^{(d)},}
 and
$\delta I^{(N)}$ is a boundary contribution given by
\eqn\boundef{\delta I^{(N)} = \int_0^\infty d\beta {d\over d\beta}\tr (-1)^F 
e^{-\beta H_N} .}

  The boundary term 
was found to arise  in the $N=2$ case from the
region of moduli space in which the two particles are separated and 
non-interacting 
\yia\sethia.  This region is described by quantum mechanics of two free 
particles
on the orbifold space $R^{d-1}/S_2$.
In \greengutc\ it was assumed that, in the $d=10$ case,  for general $N$ the 
boundary term comes from
the obvious generalization of this region to the regions 
of moduli space in which $m$ ($>1$) D-particles of charge $k$
are separated, where $N=km$.  A key assumption is that the separated 
charge-$k$ 
bound states also  behave as free  particles. This region is then described by 
quantum 
mechanics on 
$R^{9(m-1)}/S_m$ and summing over the all values of $m$ leads to the 
expression  in this case of 
\eqn\boundterm{\delta I^{(N)} = -\sum_{m|N \atop m>1}  {1 \over m^2}.}
The Witten index follows once $I^{(N)}_{bulk}$ is evaluated.

By making use of \bulkdef\ the expression for $I^{(N)}_{bulk}$  can be 
identified with the coefficient  of the $N$-instanton contribution to the 
expansion of
certain protected terms in the type IIB effective action.  These
terms include  \greengutd\gv\greengute, 
\eqn\termss{\int d^{10}x \sqrt{g} e^{-\phi/2}(f_4(\rho,\bar \rho) R^4  + 
f_{16}(\rho,\bar \rho) \lambda^{16} + \cdots),}
where  
$R^4$ denotes a specific contraction of  four Riemann curvatures,
$\lambda^{16}$ is a specific sixteen-fermion term (where $\lambda$ is 
the spin-1/2 Weyl fermion of the IIB theory) and the terms indicated
by dots are other terms of the same dimension that are related by 
supersymmetry 
\greengute\kehab. Such terms include $\psi^2 \psi^{*2}$,  $f_4 G^2 G^{*2}$,   
$f_8 G^8$ and many others, where $\psi_\mu$ is  the Weyl gravitino
 and $G$ is a 
complex combination of the Ramond--Ramond (\RR)  and the 
Neveu-Schwarz--Neveu-Schwarz (\NSNS) antisymmetric tensor field 
strengths\foot{For related work on D-instanton effects  
  in toroidally compactified 
type II 
theories see   \kiritsisb\kiritsisc\piolinea\ and in  type I theories see
\bacha\bachb\kiritsisa. 
 $SL(2,Z)$-invariant expressions for higher-dimensional terms in the type IIB 
effective
action have also been proposed in  \russoa\russob\kehaa\kehab\berkova.}. The 
functions $f_4$, $f_{16}$ and the other coefficient functions 
are nonholomorphic modular forms that depend on  
the complex scalar field 
\eqn\rhodef{\rho = c^{(0)} + i e^{-\phi},}
 where
$c^{(0)}$ is the \RR\ scalar and $\phi$ is the dilaton\foot{In the 
earlier papers \greengutd\greengute\ the function 
$f_4$ was denoted $f$ -- a more uniform notation is adopted here.}.
These coefficients  are  generalized (nonholomorphic) Eisenstein series
and are given by
\eqn\eisendef{f_p = \rho_2^{3/2} \sum_{m,n \ne 0} (m + n \bar \rho)^{p- 11/2}
(m+ n \rho)^{-p + 5/2},}
which  transforms under modular transformations as a form
 of holomorphic weight $( p - 4)$ and antiholomorphic weight $-p + 4$ which
we will write as 
\eqn\weights{{\rm Weight}\ f_p = ( p - 4, -p + 4) .}
These functions are related to each other by the action of a covariant 
derivative,
\eqn\ovrel{f_p = 2^p {\Gamma(-5/2)\over \Gamma(p-5/2)}\big(\rho_2 {\cal 
D}\big)^p f_0 ,}
 where
\eqn\ddef{{\cal D} =  {i \partial \over \partial \rho} + {d \over 2\rho_2},}
acting on a $(d, \bar d)$-form converts it
 into  a $(d+2, \bar d)$-form.

  The expansion of the coefficient functions for 
small coupling, $e^\phi \to 0$, gives perturbative  tree-level and
one-loop contributions together with an infinite number of D-instanton (and
anti D-instanton) terms.   The absence of higher-order perturbative
 corrections  
is presumably  related to the fact that  the terms in \termss\ are given by 
integrals over
half the on-shell superspace. An indirect argument for such a 
non-renormalization theorem has been advanced in  \berkov.  The expansion of 
$f_p$  can be written in the form \genexp,
\eqn\expadef{e^{-\phi/2} f_p = 2\zeta(3) e^{-2\phi} + {2\pi^2 \over 3} c_p + 
\sum_{N=1}^\infty 
\calG^{(10)}_{N, p},}
where \kehaa
\eqn\cpdef{c_p = (-1)^p {\pi \over 4} {1 \over \Gamma(-5/2+ p)
\Gamma(11/2 -p)}.}
The first two terms in \expadef\ have the interpretation of the 
 tree-level and 
one-loop string terms while the instanton and anti-instanton terms
 are contained 
  in the asymptotic series,
\eqn\znpddef{\eqalign{\calG^{(10)}_{N,p} =& (8\pi)^{1/2}  \left(\sum_{N|m} 
{1\over m^2}\right) (2\pi N\rho_2)^{1/2} \cr &
\times \left(\sum_{r=4-p}^\infty{c_{-p, r}\over (2\pi N\rho_2)^r}e^{2\pi i 
N\rho} + 
\sum_{r=p-4}^\infty {c_{p,r}\over (2\pi N\rho_2)^r} e^{-2\pi i N\bar \rho} 
\right),\cr}}
where
\eqn\ckrdef{c_{p,r} = {(-1)^p \over 2^r (r-p+4)!} {\Gamma(3/2) \over 
\Gamma(p-5/2)} {\Gamma (r-1/2)\over \Gamma(-r - 1/2)}.}

The expression \znpddef\ is somewhat formal since the series multiplying each 
(anti)instanton contribution are asymptotic 
 expansions of Bessel functions for large argument.

\ From \znpddef\ we see that  
the leading contribution to the $N$ D-instanton contribution in  
$\calG^{(10)}_{N, p}(\rho,\bar \rho)$ is  
\eqn\coeform{\calG^{(10)}_{N,p} \sim   Z^{(10)}_N  (2\pi S_N)^{-7/2 + p} 
e^{-2\pi  (S_N + 
iN c^{(0)})} \left(1+ o(e^{\phi})\right),}
where the $N$ D-instanton action
is given by $S_N = N e^{-\phi}$.  This expression is of the form \coefform\
with $a_{10} =   -7/2$, $\chi $ identified with $c^{(0)}$ and with
 \eqn\mudef{Z_N^{(10)} = \sum_{m|N} {1 \over m^2},}
which has been normalized so that $Z^{(10)}_1 =1$. 

The 
 power  $a_{10}=-7/2$ in \coefform\ arises from the combination of ten
 bosonic zero modes
 (each contributing $S_N^{1/4}$) and
 sixteen  fermionic zero modes  (each contributing $S_N^{-3/8}$).   The number 
$p$ is the number of external fields in the linearized approximation to the 
interaction term ($p=4$ for $R^4$, $p=16$ for $\lambda^{16}$, etc.).     

We have thus determined the   factor that is to be 
identified with the partition function of the SU(N) zero-dimensional matrix
mode.
Combining \mudef\ with  \bulkdef\ and \boundterm\ gives the result for
 the Witten 
index, \indform, of the
$d=10$ case,
\eqn\witind{I^{(N)} = 1,}
which indicates the presence of at least one bound state for each value of $N$.

\newsec{$d=6$ and heterotic-type II duality}

In this section  we wish to consider type IIA compactified to six dimensions 
 on 
$K3$, which is equivalent, 
via strong coupling duality, to the heterotic string on $T^4$. This
 theory has   
two eight-component supercharges. At special points in the 
heterotic moduli space 
${\cal M}_{4,20}$ additional massless states appear leading to a perturbative 
enhancement of the gauge symmetry.  On the IIA side these are associated with 
D2-branes wrapping  a vanishing $S^2$ in
 the K3. The absence of bound states of multiply-wrapped D2-branes is 
demanded by the absence of an infinite tower of such massless gauge states on 
the heterotic side.
A T-duality transformation (in the euclidean time direction), maps the D2-brane 
of IIA to the euclidean world-sheet of a D-string in IIB. This
 is  a D-instanton 
\ from the six-dimensional point of view.    

Since the $S^2$ is chosen to be a supersymmetric cycle such an instanton 
preserves half of the  supersymmetries \becker. The eight broken 
supersymmetries generate fermionic collective coordinates
which have to be soaked up by external sources in order to give non-zero 
correlation functions in an 
instanton background, thereby generating new interaction vertices. The 
simplest 
such term will be an eight-fermion term.  Among the other terms 
related to this 
by supersymmetry is 
the  four-derivative term $ \partial_\mu\phi\partial_\nu \phi 
\partial^\mu\partial^\nu \phi$. In 
the background of 
$N$  D-instantons such vertices are weighted with a factor $\exp(-S_{N})$
 where 
the $N$ D-instanton action is given by
\eqn\instacts{S_{N}= N e^{-\phi_6} {\rm Vol}(S^2),}
where $\phi_6$ is the $d=6$ dilaton.
We will obtain the expression for   $Z^{(6)}_N$ 
\ from the form 
of these four-derivative interactions in the effective action.

  This is seen by considering 
one-loop threshold corrections to the  heterotic string on $T^6 = T^4 
\times T^2$ \antoniadis.    One of these is of the form 
 $F_1 R^2$ term  where $F_1$ is a function of the vector multiplet moduli  
while the other is   $\tilde F_1 \partial_\mu\phi\partial_\nu 
\phi 
\partial^\mu\partial^\nu \phi$  where $\tilde F_1$  depends on 
the hypermultiplet moduli only. 
Since 
the heterotic dilaton lies in a vector 
multiplet the heterotic one-loop calculation of the threshold function 
$\tilde{F}_1$ is exact.  In this 
manner the form of $\tilde F_1$ may be determined  in type IIA on $K3 \times 
T^2$, which may then be  related by T-duality in one of the $T^2$ directions 
to 
type IIB.  This duality, which  exchanges the K\"ahler modulus and 
complex modulus of $T^2$,  exchanges a two-brane wrapped on a two-cycle for 
the 
world-sheet of a D-string wrapped on the same cycle.  
The six-dimensional four-derivative terms can then be found by taking 
the large 
volume limit of the type IIB $T^2$. 

Thus, following \antoniadis,  we may write the six-dimensional one-loop result 
for $\tilde{F}_1$ in
 terms of 
type IIB variables and extract  the various perturbative and nonperturbative 
contributions  by expanding in the appropriate couplings. 
   The expression for 
$\tilde F_1$ is 
given in eq. (6.2) of \antoniadis,
\eqn\antonres{\eqalign{\tilde{F}_1= & 8\pi+ 2 e^{-\phi_6} \sum_{N\neq 0,
 q^i\neq 
0}{1\over 
|m|} C\left({q^tLq\over 2}\right) ({q^t(M+L)q})^{1\over 2}\cr
& \qquad  K_1\left(2\pi |N|  
\left({q^t(M+L)q\over 2}\right)^{1\over 2}e^{-\phi_6}\right)e^{2\pi i N Y_i 
q^i},\cr}}
The matrix $M$ parameterizes the moduli space $O(20,4,Z)\backslash 
O(4,20,R)/O(4,R)\times O(20,R)$, $L$ is the metric on the signature $(20,4)$  
Narain lattice    and $q^i$ are integer charges that parameterize the lattice  
momenta ($p_l$ and $p_r$) that satisfy,   
\eqn\norms{p_l^2={1\over 2} q^t(M+L)q,\quad p_r^2={1\over 2}q^t(M-L)q.}
The level matching and mass-shell conditions for  perturbative heterotic
 states are 
\eqn\masscon{{1\over 2}(p_l^2-p_r^2) = N_l-N_r+1,\qquad
m^2 = {1\over 2}(p_l^2+p_r^2)+N_l+N_r-1. }
The degeneracy factor $C(k)$ in \antonres\ is defined by 
$\eta^{-24}(\tau)=\sum_{k>-2}C(k) 
e^{-k\tau}$ ($\eta$ is the Dedekind function) and $\phi_6$ is the 
six-dimensional type IIB
dilaton.  The Wilson lines, $Y_i$, of the heterotic string correspond to
 \RR\  
fields dimensionally reduced on $K3$  on the type II side.

 In \antoniadis\ the expression \antonres\ was used to determine the 
non-perturbative effects in the decompactified limit in which the
 volume of the 
$K3$ is infinite, which confirms the form of the $d=10$ D-instanton terms 
described in section 2.  Here, we wish to use \antonres\ to extract the $d=6$  
D-instanton corrections coming from euclidean D-branes 
wrapping 
the 2-cycles of $K3$.
Using the mass-shell 
condition it 
is easy to see from the heterotic side that the extra state with 
vanishing mass 
has a charge satisfying $q^tLq/2=-1$ and a mass given by  
\eqn\massze{\mu^2 =  {q^t(M+L)q\over 2}.}
 To isolate the instanton effects on the type IIB side we tune the moduli
 $M$ 
in such a way that $\mu \to 0 $  for a specific charge vector $q_i$, such that 
$\mu e^{-\phi_6}$ is fixed.  
 The degeneracy of this state is $C=1$.
Expanding the Bessel function $K_1$ for large  $\mu e^{-\phi_6}$ gives 
the  leading 
nonperturbative contribution,
\eqn\sixres{\eqalign{\tilde{F}_1&\sim \sum_{N \ne 0} e^{-\phi_6/2} 
{\mu^{1/2}\over 
|N|^{3/2}} e^{-2\pi  |N| \mu e^{-\phi_6}}e^{2\pi i  N 
\int_{S^2}c^{(2)}} +c.c.\cr
&= \sum_{N \ne 0} {1\over N^2} (S_N)^{1/2}  e^{-2\pi S_N+2\pi  i N 
\int_{S^2}c^{(2)}} +c.c.},}
where we have kept only the first term in an asymptotic series of  perturbative 
fluctuations around the D-instanton 
background, $c^{(2)}$ is the two-form \RR\ potential that couples to the wrapped 
$D$-string world-volume   and  we have used  $\mu = 
Vol(S^2)/\alpha^\prime$ so that  $S_{N}= 
|N| \mu 
e^{-\phi_6}$  is the instanton action.  This expression again has the form 
\coefform\ with $a_6=-3/2$ and $p=2$, where $\chi$ is  identified with
 $\int c^{(2)}$.
With this identification we determine,
\eqn\ressix{Z_N^{(6)}={1\over N^2},}
a result which was also derived in \mooreetal\ using apparently 
different arguments.

This result  implies that the bulk term in the Witten vertex
 is $I^{(N)}_{bulk} 
= 1/N^2$.  In the absence of multi 
D-particle threshold bound states (multiply-wrapped D2-branes)  the boundary 
contribution to the  index is that of $N$ free particles, namely, $\delta 
I^{(N)} = -1/N^2$.  This is consistent with the expectation that the Witten 
index  $I^{(N)}_{bulk} + \delta I^{(N)}$ should vanish in the $d=6$ case. 

\newsec{$d=4$  and the conifold}
\seclab\conifold
We turn now to consider the type II string theories compactified on a 
Calabi--Yau threefold 
near a conifold singularity. In the simplest case   a 
nontrivial three cycle $\gamma$ with period 
\eqn\period{z= \int_\gamma \Omega}
 vanishes in the conifold limit.
 As pointed out by  Strominger \stromingera\ the singularity in the 
 vector multiplet moduli space
of the classical  IIB theory  
is
interpreted   in the low energy  quantum theory 
as  the one-loop effect of a light hypermultiplet produced 
by a D3-brane wrapping the cycle $\gamma$.
In contrast to the vector multiplet moduli space the hypermultiplet moduli 
space 
can receive both perturbative and nonperturbative corrections.
 It was suggested in \becker\ that for IIA on the same Calabi--Yau space 
euclidean 
D2-branes wrapping $\gamma$ lead to large instanton effects which smooth out
 the 
classical singularity of the hypermultiplet moduli space. Following field 
theoretical arguments given in \greene\  a corrected metric  on the moduli
 space 
near the conifold was
 derived in \ooguri. The metric was determined in the limit $z\to 0$ keeping 
$|z|/\lambda$  fixed (where $\lambda$ is the string coupling),  in which the 
details of 
how the conifold is embedded in 
the Calabi--Yau do not play any r\^ole (neither does the fact that the 
hypermultiplets 
parameterize a quaternionic instead of 
a hyperk\"ahler geometry).  The expression for the metric
in \ooguri\ is given   (in the string frame) by,
\eqn\oogurimet{ ds^2= V^{-1}\big( dt - A_x dx - {1\over \lambda}A_zd\bar{z}- 
{1\over \lambda}A_{\bar{z}} dz\big)^2+ V\big( dx^2+ {1\over 
\lambda^2}dzd\bar{z}\big),}

The four 
scalars in the  hypermultiplet comprise the complex field $z$ associated 
with the complex structure deformation parameterizing the conifold limit and 
$x,t$ which are  the  reduced \RR\ three-forms corresponding to  the 
elements of the 
cohomology  associated with the cycle $\gamma$ and the dual cycle respectively.
The scalar potential $V$  in \oogurimet\ is  given by\foot{In order to avoid 
notational confusion we are using the symbol $\Lambda$ instead of $\mu$ to 
represent the renormalization scale.} 
\eqn\vresult{V= {1\over 4\pi} \ln\left({\Lambda^2 \over |z|^2}\right) 
+{1\over \pi}
\sum_{N> 0}
\cos(2\pi N x) K_0\big( 2\pi N |z|/\lambda \big),}
and the vector potential $A$ is determined by $\nabla V= \nabla 
\times A$ to be, 
\eqn\aresult{A_x= -{1\over 2\pi}\theta,\quad 
A_z=A_{\bar{z}}= {1\over \pi}\sum_{N>0}\sin(2\pi Nx) K_{1}\big(2\pi 
 |N z|/\lambda \big).}
The instanton terms again become apparent by expanding the Bessel function 
$K_0$ 
for large values of $|z|/\lambda$, 
\eqn\expbess{V\sim {1\over 4\pi}  \ln\left({\Lambda^2 e^{2\Phi}\over 
\lambda^2}\right) + \sqrt{2\over \pi}\sum_{N >0} 
\left(2\pi{N e^{-\Phi}}\right)^{-1/2} e^{- 2\pi {N e^{-\Phi}}} 
\cos(2\pi Nx)(1 + o(1/Ne^{-\Phi})),}
where we have set $z = \lambda e^{-\Phi + i\theta}$ so that $|z|/\lambda = 
e^{-\Phi}$ and
\eqn\phimet{dz d\bar z = \lambda^2 e^{-2\Phi} (d\Phi + id\theta)(d\Phi - i 
d\theta).}

\ From this it follows that the non-perturbative contribution to the $d\Phi 
d\Phi$ component of the metric has the form
\eqn\metriform{\sum_{N\neq 0}\calG^{(4)}_{N,\Phi\Phi}  d\Phi d\Phi  ,}
where $\calG^{(4)}_{N,\Phi\Phi}$ 
has a leading $N$-instanton contribution of the 
form \expbess,  
\eqn\zinst{\calG^{(4)}_{N,\Phi\Phi} \sim {1\over 2 \pi} {1\over N^2} 
S_N^{3/2}e^{2\pi i N x} 
e^{-2\pi  S_N}+cc,}
and where the action for a charge $N$ instanton is given by
\eqn\instacta{ S_N= |N|e^{-\Phi}.}
In writing \zinst\ we have kept only the leading contribution to the 
$N$-instanton term, which does not get contributions from the expansion of the 
expression for $A_z$ in \aresult.
The expression \zinst\ is again of the form \coefform\ with $\chi$ identified 
with $x$, $p=2$, $a_4 =-1/2$ and  
with 
\eqn\dfourres{ Z_N^{(4)}  =  {1\over N^2}.}
 As with the $d=6$ case this expression is consistent with the 
absence of D-particle bound states (and agrees with the result in \mooreetal). 

In identifying   the measure \zinst\  it was important that we used the metric 
for the 
fluctuations of $\Phi$ rather than of the field $z$. 
 These metrics differ by a 
factor 
of  $\lambda^2 e^{-2\Phi}$.    One check that this choice of normalization is 
appropriate for extracting the form of $Z_N^{(4)}$ is to emulate the way in 
which the 
instanton measure can be evaluated in field theory by considering the 
contributions of perturbative fluctuations around the instanton background to 
correlation functions.  

\subsec{Perturbative fluctuations around a stringy D-instanton}
The appearance of D-instanton induced terms in the effective 
action can be seen in   string perturbation theory around an 
instanton background following  the same kind of arguments as were made in 
\greengutd\ 
for the ten-dimensional theory.  To lowest order in a perturbative 
expansion the world-sheet consists of a number of disconnected disks with 
closed-string vertex operator insertions and with D-instanton boundary 
conditions which can be implemented by 
constructing an appropriate boundary state.  The D-instanton preserves half 
the  space-time supersymmetry so that a 
combination of the left-moving and right-moving space-time supersymmetry 
charges 
annihilate the boundary state.  Applying $n$ of
 the broken supersymmetries 
to the boundary state generates a boundary state,  $|B \rangle_n$, where $n$ 
denotes the number of fermionic zero modes, which correspond to fermionic 
open-string ground states attached to the boundary.  Nonvanishing correlation 
functions arise when sufficient of these fermionic modes are integrated 
over --- sixteen   for the IIB theory in $d=10$ considered in 
\greengutd\ (and section 2), eight in the $d=6$ case in section 3 and four in 
the 
$d=4$ case in section 4.  
 
Here we will only consider the $d=4$ case where the instantons are  euclidean 
D2-branes wrapped on a three-cycle of a Calabi--Yau threefold.
The implementation of the  boundary conditions
in this case, as well as 
the construction and properties of the associated boundary states
 were discussed
 in 
detail in \oogurioz.
Type IIA compactified on a Calabi--Yau manifold has eight real supersymmetries 
and a euclidean 
D2-brane wrapped on a supersymmetric three-cycle preserves four of these.
 Hence  there are four fermionic zero modes and this leads, for example, 
 to a \lq 'tHooft' four-fermion 
interaction vertex 
\becker.  The leading contribution to this interaction comes from a 
 world-sheet 
consisting of four disconnected disks with a fermion vertex operator  attached 
to the 
interior of each and one fermionic zero mode (open string) attached to each 
boundary.
This four-fermion vertex is related 
by supersymmetry to the metric for the hypermultiplets that we have been 
considering. 
The leading instanton contribution  to this metric
 comes from  a world-sheet consisting of two disconnected disks with a 
vertex for a 
hypermultiplet modulus inserted to the interior of each  and two 
fermionic zero 
modes 
attached to each boundary. 

The vertex for the 
hypermultiplet modulus $\Phi^i$ is given by,
\eqn\closone{V_\Phi^i = e^{-\phi}e^{-{\bar{\phi}}} \Phi^i_{q,\bar{q}}e^{ikX},}
where $q$ and $\bar{q}$ denote the left and rightmoving $U(1)$ charges of the 
field $\Phi^i$ of the internal $c=9$ $N=2$ super-conformal field theory which
 is 
associated with the 
compactification on the Calabi--Yau manifold. 
For type IIA compactified on a Calabi--Yau  the 
hypermultiplets parameterize the complex-structure moduli space and the 
scalars 
in the \NSNS\ sector are associated 
with elements of the cohomology $H^{2,1}$ and $H^{1,2}$ which are given by 
$(c,c)$ primary fields $\Phi^i_{1,1}$ and $(a,a)$ fields 
$\Phi^{\bar{i}}_{-1,-1}$ respectively.  The \RR\ vertices are related to these 
by spectral flow.

In \oogurioz\ it was shown that there are two possible boundary conditions 
called A and B (connected to the two topological twists of $N=2$ SCFT
 \wittentop\ 
which correspond to branes wrapping middle and even dimensional cycles 
respectively.
Hence the $A$ boundary conditions are relevant in our case and the
 supersymmetry 
charges are given in the $-1/2$ picture by
\eqn\susyone{\eqalign{Q^a_{-1/2\;\pm}&= e^{-1/2 \phi} S^a e^{-i{\sqrt{3}\over 
2}H}\pm i  e^{-1/2 \bar{\phi}} \bar{S}^a e^{i{\sqrt{3}\over 2}\bar{H}},\cr
Q^{\dot{a}}_{-1/2\;\pm}&= e^{-1/2 \phi} S^{\dot{a}} e^{i{\sqrt{3}\over 2}H}\pm
 i e^{-1/2 \bar{\phi}} \bar{S}^{\dot{a}} e^{-i{\sqrt{3}\over 2}\bar{H}} }}
and in the $+1/2$ picture by
\eqn\susytwo {\eqalign{Q^a_{1/2\;\pm}&= e^{1/2 \phi}(\gamma_\mu  S)^a \partial 
X^\mu  e^{-i{\sqrt{3}\over 2}H}\pm i e^{1/2 \bar{\phi}}(\gamma_\mu \bar{ S})^a 
\bar{\partial} X^\mu  e^{i{\sqrt{3}\over 2}\bar{H}},\cr  
Q^{\dot{a}}_{1/2\;\pm}&= e^{1/2 \phi} (\gamma_\mu  S)^{\dot{a}}\partial X^\mu 
e^{-i{\sqrt{3}\over 2}H}\pm i e^{1/2\bar{ \phi}} (\gamma_\mu  
\bar{S})^{\dot{a}}\bar{\partial} X^\mu e^{i{\sqrt{3}\over 2}\bar{H}}.}}

Here the unbarred fields denote leftmovers and the barred fields denote 
rightmovers, $\phi$ denotes the bosonized superghost, $S^a$ and $S^{\dot{a}}$ 
are  $SO(4)$ spin 
fields of opposite chirality,
 and $H$ is the free boson associated with the  $U(1)$ current of the 
internal $c=9$ SCFT. The $A$ boundary condition
 enforces $Q_+ | B\rangle=0$ and 
$Q_-$ are the vertex
 operators 
for the fermionic collective coordinates of the D-instanton.
The disk amplitude is then given by inserting one scalar vertex and two 
vertices for supercharges of the broken supersymmetries (such amplitudes were 
evaluated in the ten-dimensional  case in \greengutd\guta),
\eqn\nsdisk{\eqalign{& \phi_i \epsilon_1^a\epsilon_2^{\dot{a}}\left\langle 
c\bar{c} 
e^{-\phi}e^{-{\bar{\phi}}} \Phi^i_{q,\bar{q}}e^{ikX}(z)\;c Q_-^a(x_1)\;\int
 dx_2 
Q_-^{\dot{a}}(x_2)\right\rangle \cr
& = \bar{\epsilon}_1 \gamma^\mu \epsilon_2 \partial_\mu  
\phi_i \ {N\over \lambda} \langle \Phi^i \rangle , \cr}}
where $\phi^i$ is the (on-shell) wave function for the field $\Phi^i$.  
One supersymmetry charge is in the $1/2$ and one in the $-1/2$ 
picture in order for the total ghost number on the disk to add up to $-2$.
The notation  $\langle \Phi^i \rangle$ denotes the expectation value
 on the disk 
of the 
\RR\  field $\Phi^i_{-1/2,-1/2}$ which is associated to $\Phi_{1,1}$ by 
spectral 
flow and is the same as the  topological amplitude derived in \oogurioz. There 
it was shown that $
\langle \Phi^i \rangle$ is independent of the K\"ahler moduli   and is given by
\eqn\oogres{\langle \Phi^i \rangle = \int_\gamma \omega^i= D_i\int_\gamma 
\Omega, } 
where $D$ denotes the covariant derivative on the vacuum line bundle over the 
moduli space of the $N=2$ SCFT \ooguri.

The lowest-order term in the correction to the metric comes from a
 configuration 
of two disks.  One of these has a vertex operator for the modulus $\phi^i$ 
attached and the other has the vertex operator for the 
complex conjugate $\phi^{\bar{j}}$.   After integration over the fermionic 
zero 
modes and summation over the 
instanton sectors the expression reduces to 
\eqn\sclardi{\eqalign{\sum_{N \ne 0}\int d^4\epsilon 
\langle V_{\phi^i} \epsilon 
Q\epsilon Q\rangle \; \langle V_{\phi^{\bar{j}}} \epsilon Q\epsilon Q\rangle 
\calZ^{(4)}_N &= \partial_\mu \phi^i \partial^\mu \phi^{\bar{j}} 
\sum_{N \ne 0} 
{N^2\over 
\lambda^2} \calZ^{(4)}_N D_i
\int_\gamma \Omega\;  D_{\bar{j}}\int_\gamma \bar{\Omega}, }}
where $\calZ^{(4)}_N$ is to be identified with the $N$-instanton 
measure -- it is a  factor that does not depend on the 
particular process being considered.

We now specialize to the case where the vertices $\phi_i,\phi_{\bar{j}}$ 
are the moduli $z,\bar{z}$ respectively. In the conifold limit  $D_z 
\int_\gamma 
\Omega=1$ and it follows that the instanton correction to the metric 
is  given by
\eqn\instmeta{\eqalign{\sum_{N\ne 0} {N^2\over 
\lambda^2}\calZ^{(4)}_N  dz d{\bar{z}} &= {\sum_{N\ne 0}
 (Ne^{-\Phi})^2}\calZ^{(4)}_N  \left(d\Phi d\Phi+d\theta d\theta\right)\cr
 &=\sum_{N\ne 0} 
\calG^{(4)}_{N, 
\Phi\Phi} \big(d\Phi d\Phi+d\theta d\theta\big) .}}
In order for \instmeta\ to   agree with  \metriform,  $\calG^{(4)}_{N, \Phi\Phi}$ must be identified 
with  \zinst. In the analysis of the instanton induced corrections
 of the metric $\calZ^{(4)}_N$ is interpreted as the instanton measure 
which is independent of the process considered. On the other hand the disk 
amplitudes \nsdisk\ depend on the fluctuating fields. In the case of the
 fluctuations of the field  $z$ the disk amplitude  \nsdisk\ is proportional
 to $N/\lambda$ which is not equal to the instanton action $S_N$. In order
 to bring  to bring \instmeta\ into the general form \coefform\ with
 $a_4=-1/2$ and $p=2$ it is necessary to consider the fluctuations of the
 fields $\phi$ and $\theta$ instead of $z,\bar{z}$ which can be accomplished
 by the change of variables \phimet.

The leading instanton correction for to the metric for  the \RR\ 
scalar 
$x$ is given by $V dx dx$ and  is also reproduced by the two-disk process. The 
\RR\ fields have a different dependence on the string coupling (the one-point 
function is $\langle V_x \rangle = N$ in this case) so it is important to 
remember that the canonically normalized perturbative field is $\hat x=x 
e^{-\Phi}$.  The  metric is then of the form 
\eqn\xmetri{{1\over N^2} (S_N)^2V d\hat x d\hat x,}
which may be expressed as  \coefform\ with $a_4=-1/2$ and $p=2$.

All the other corrections in 
\oogurimet\ are subleading in the coupling constant expansion and should 
correspond to more complicated diagrams in the instanton background which will 
not be discussed here.

It would be interesting to derive the form of the metric \oogurimet\ from the 
duality of type II on Calabi--Yau and the heterotic string on $K3\times T^2$.
As in the discussion of threshold corrections in the $d=6$ case,
since the heterotic dilaton 
lies in a vector 
multiplet the nonperturbative effects on the type II side must be reproduced
 by  tree level effects on the heterotic side (taking into account all orders 
in $\alpha^\prime$). Unfortunately, relatively little 
is known about the structure of such hypermultiplet moduli spaces from 
the heterotic point of view (for a
 recent  review  see \dualhet).

\newsec{Discussion}

We have considered the contributions of D-instantons to a variety of protected 
interactions in $d=4,6$ and $10$ dimensions.  There appear to be no 
examples of such 
instanton effects when $d=3$, which is consistent with the expected 
vanishing of
 $Z_N^{(3)}$.
The interaction terms considered are ones that are protected by supersymmetry
 and in 
which only multiply-charged single D-instantons contribute  since
 the contributions 
of separated multi-instantons carry extra fermionic zero modes and
 therefore vanish. 
   We are thus able   to isolate the instanton partition function,
 $Z_N^{(d)}$, 
which also only gets contributions from single instantons.  

In all the cases considered in this paper the basic structure of the 
$N$-instanton calculation in $d$ dimensions  follows from a one-loop calculation 
$R^{d-1} \times S^1$ with a circulating D-particle of the appropriate type.  
Using  T-duality the charge-$N$ D-instanton corresponds to the contribution of  
an $N$-fold winding of the euclidean world-line of the D-particle around the 
$S^1$.  An expression of the form \coefform\ follows in a simple manner in each 
case, as is summarized in the appendix.  It is important to note
 that in all 
cases there is an infinite tower of multiply-charged D-instantons.

The correspondence, via euclidean T-duality, between D-instantons and
 D-particles\foot{Some apects of this relation were discussed 
in \lustb\bergb.} 
makes the connection with the Witten index clear.  In  $d=4$ and 
$d=6$  the D-particle is the unique normalizable state --- described by a 
D3-brane 
wrapped around a three-cycle of a Calabi--Yau space (for $d=4$) or   a 
D2-brane 
wrapped around a two-cycle of K3 (for $d=6$).  In the $d=10$ case the 
D-instanton is 
associated with a pair of integers 
 which are identified, after T-duality, with 
the  
charge of a  D-particle threshold bound state and with  the winding number of 
its 
world-line.

The power of the instanton action $(S_N)^{a_d}$ which enters the  measure 
 in \coefform\ takes the values $a_d=-7/2,-3/2$ and $-1/2$ for 
$d=10,6$ and $4$, 
repectively. This factor should emerge from the  
Jacobian for the change of variables from  zero modes to  collective 
coordinates.  Although we have not done that calculation explicitly, these 
values of $a_d$ are consistent with attributing a factor of 
$S_N^{1/4}$ to 
each bosonic collective coordinate and $S_N^{-3/8}$ for each fermionic 
collective coordinate.
   
Having the exact partition function for all $N$ might be of significance
 for the 
zero-dimensional matrix model \japanese\ and other applications of large-$N$ 
 supersymmetric  Yang--Mills theory.   In that context it is perhaps notable 
that the 
$d=10$ expression \mures\ does {\it not} have a well-defined large-$N$ limit.
  The 
special interactions considered in this paper (such as the 
$R^4$ term) are interpreted in the Yang--Mills matrix model as a special class 
of local  gauge-invariant correlation functions.  These can be expressed as
 correlations of small Wilson loops 
that are associated with punctures, or vertex operators, in the string 
picture.

\vskip 0.4cm

\noindent {\bf Acknowledgements}
\medskip
\noindent M.B. Green  wishes to
 acknowledge useful conversations with Steve Shenker, Andy Strominger, Hiroshi 
Ooguri, Wolfgang Lerche, Tom Banks and other participants of the Workshop on "Dualities in String Theory"  at the Institute for Theoretical Physics, Santa Barbara.
M. Gutperle thanks Ori Ganor, Yuji Satoh, Eric Sharpe, Rikard von Unge and 
Dan Waldram for 
conversations and gratefully  acknowledges the partial support  by DOE grant  
DE-FG02-91ER40671, NSF grant
PHY-9157482 and a James. S. McDonnell grant 91-48.

\appendix{A}{}

For all values of $d$  considered in this paper the instanton-induced terms in 
the  effective action may be simply summarized by expressing them in terms
 of a 
one-loop  Feynman diagram with $p$ external states in the T-dual theory on 
$R^{d-1} \times 
S^1$ (this is the spirit in which the $d=4$ case was discussed in 
\shenkerseiberg). Integration over the 
fermionic modes gives rise to the interaction $\calP(\Psi)$ leaving a one-loop 
amplitude for a scalar field theory that determines the coefficients ${\cal 
F}_\calP^{(d)}$ which are functions of the moduli.  This amplitude is given by 
 \eqn\loopa{A_{\cal P}= R \pi^{-d/2}\sum_n \int d^{d-1}{\bf p}
 \int {dt\over t} 
t^p 
e^{-t\big( 
p^2 
+{(n-x)^2\over R^2} +\mu^2\big)},}
where $R$ is the radius of the (euclidean) circle,  $n$ is the Kaluza--Klein 
charge and the factor $t^p$ originates from integrating over the proper
 times of 
the $p$ vertex operators around the loop.  The shift in the integer
 momentum is 
due to a non-zero \RR\ Wilson line in the 
compact direction.

Integration over the loop momentum ${\bf p}$ and a Poisson resummation with
 respect to $n$ gives.
\eqn\loopb{A_{\cal P} = R^2  \sum_m \int {dt\over t} t^{k-d/2}
 e^{- {\pi^2R^2m^2\over t}- t \mu^2}e^{2\pi i m x},}
where $m$ is the winding number of the world-line around the compact
 dimension.  
In this form the ultraviolet divergence of the loop amplitude arises in
 a single 
zero-winding number ($m=0$) term.   The terms with nonzero $m$ give the 
instanton corrections that we are interested in here.

 The cases under consideration in this appendix are those 
with 
 $k- d/2=-1$.  These one loop amplitudes contain vertex operator
 insertion which are 'maximal' in the sense that each vertex operator
 absorbs four fermionic zero modes and corresponds to a two derivative 
term on a bosonic field --- the 
four vertices of linearized $R^4$ in $d=10$, the two vertices
of  $\partial^2\phi\partial^2\phi$ in $d=6$ and the one 
 $\partial^2\phi$ vertex 
in $d=4$. Note that in the case $d=4$ such a term is related to the
 corrections of the metric discussed in section 4 by an integration
 by parts as will be  seen explicitly at the end of this appendix.

 After a 
change of variables $t \to 1/t $ these amplitudes are given by 
\eqn\loopc{\eqalign{A_{\cal P}&= R^2  \sum_m \int dt
 \exp({- t\pi^2 R^2m^2-{\mu^2\over t}})e^{2\pi i m x}\cr
   &= \pi^{-2}\sum_m {1\over m^2} \int dt \exp({- t-{(2\pi 
m R\mu)^2\over 4 t}})e^{2\pi i m x}\cr
&= \pi^{ -2}\sum_m {1\over m^2} (2\pi m R\mu) K_{-1}(2\pi
 |m| R\mu) e^{2\pi i m x}}}
 In general, the mass $\mu$ of the D-particle 
is a function of the moduli of the 
form  $\mu=
 f(z^i)e^{-\phi_d}$ where $\phi_d$ is the $d$-dimensional dilaton.    This 
transforms to $\phi'_d$ under T-duality where, 
\eqn\instacta{2\pi m R\mu=2\pi m R f(z)e^{-\phi_d}= 2\pi m 
f(z)e^{-\phi_d^\prime}= S_m.}
Substituting this into \loopc\ and expanding the Bessel function for large 
argument produces a leading instanton contribution of 
\eqn\eadinst{ \sum_m {1\over m^2} (S_m)^{1/2} e^{-S_m}  e^{2\pi i m x}}
in all cases.   
This assumes that the D-particle circulating in the loop is nondegenerate,
 which is true for $d=4$ and 
$d=6$.  In the $d=10$ case there are     charge-$k$ threshold bound states  
with 
masses 
$\mu_k = k \mu$ that circulate in the loop.  The amplitude must  therefore be 
summed over $k$ as well as the winding number $m$.  Writing $N= mk$ the result 
is 
\eqn\loopd{A= \pi^{ -2}\sum_N \sum_{N|n} {1\over n^2} (S_N) K_{-1}(S_N)
 e^{2\pi i N x}. }
Hence  the different behavior of the 
instanton measure ${\cal Z}_N$ for $d=10$ compared to $d=4,6$.

The loop amplitude \loopb\ in the  $d=4$ case corresponds to the  one-point,
\eqn\loopc{
 \partial^2\Phi \sum_{N\ne 0} {1\over N^2}  | Ne^{-\Phi}|  K_{-1}(2\pi 
|N|e^{-\Phi} )
e^{2\pi i N x},}
whereas in section 4 we reviewed the correction  to the metric on the moduli space given in \ooguri\ which had the form,
\eqn\twopoint{ V e^{-2\Phi} \partial_\mu\Phi \partial^\mu\Phi,}
where $V$ is defined  in \vresult.   In fact the  nonperturbative contribution to  \twopoint\  is equal to \loopc, up to a total derivative which vanishes when integrated.  This  follows  simply from the relation among Bessel functions, 
\eqn\relk{\partial_x \big(x K_{-1}(x)\big)= - xK_0(x).}

 \listrefs
 
 \end